# Sub-µm Josephson Junctions for Superconducting Quantum Devices

J. M. Meckbach[1], *Member, IEEE*, M. Merker[1], S. J. Buehler[1], K. Ilin[1],
B. Neumeier[2], U. Kienzle[2], E. Goldobin[2], R. Kleiner[2], D. Koelle[2], M. Siegel[1]

*Abstract*—For high-performance superconducting quantum devices based on Josephson junctions (JJs) decreasing lateral sizes is of great importance. Fabrication of sub-µm JJs is challenging due to non-flat surfaces with step heights of up to several 100 nm generated during the fabrication process. We have refined a fabrication process with significantly decreased film thicknesses, resulting in almost flat surfaces at intermediate steps during the JJ definition. In combination with a mix-&-match process, combining electron-beam lithography (EBL) and conventional photolithography, we can fabricate JJs with lateral dimensions down to 0.023 µm². We propose this refined process as an alternative to the commonly used chemical-mechanical polishing (CMP) procedure. We present transport measurements of JJs at 4.2 K that yield critical-current densities in the range from 50 to $10^4$ A/cm². Our JJ process yields excellent quality parameters, $R_{sg}/R_N$ up to ~50 and $V_{gap}$ up to 2.81 mV, and also allows the fabrication of high-quality sub-µm wide *long* JJs (LJJs) for the study of Josephson vortex behavior. The developed technique can also be used for similar multilayer processes and is very promising for fabricating sub-µm JJs for quantum devices such as SQUIDs, qubits and SIS mixers.

## I. INTRODUCTION

Over the last few decades window-type (or overlap-type) Josephson junctions (JJs) have emerged as the most reproducible and controllable fabrication process for few-junctions devices such as SQUIDs [1], long Josephson junctions (LJJ) [2-5] as well as multi-junction devices like RSFQ circuits [6-8] or voltage standards [9]. To increase the performance of the devices the critical-current density $j_c$ is usually increased [10-12] while in turn the lateral dimensions are minimized. With growing complexity of the designs, an increasing number of layers become necessary. The most basic processing sequence includes deposition of the required material and etching thereof. This inherently results in a step-like topology of the chip-surface, which leads to an increase of the layer thickness in the following layers necessary in order to ensure good edge coverage. This uneven topology prohibits the use of thin resists, which limits the minimization of the lateral features sizes. To overcome these restrictions there have been many efforts including ramp-type junctions [13] or chemical-mechanical polishing (CMP). While CMP would provide almost perfectly flat surfaces, it introduces a time-consuming fabrication step and causes mechanical stress which may deteriorate the quality of the devices susceptible to strain. Additionally, small JJs themselves open up new possibilities for sub-µm devices such as SIS mixers [11], QuBits [14] and new types of meta-materials [15].

Here, we present a refined fabrication process creating effectively flat chip-surfaces at intermediate steps during fabrication. We propose this self-planarized process as an alternative to CMP. Using this technique we are able to pattern very small JJs exhibiting high quality parameters in a wide range of $j_c$. Additionally we achieve sub-µm feature sizes also in the wiring layers and the junction's periphery for the fabrication of LJJs for the investigation of fundamental vortex physics [4,5]. In Section II the fabrication of the trilayers is presented and a comparison between the so-far employed conventional process at our institute [1,2,14] and the new process will be discussed. In Section III transport measurements of short JJs will be presented along with measurements of the dependence of the maximum critical current of a LJJ on an externally applied dc-injector current [4,5]. Section IV concludes this work.

## II. FABRICATION PROCESSES

### A. Trilayer Fabrication

Our Nb/Al-AlO$_x$/Nb process is based on 2"-Si wafers which are oxidized for 4 hours at 1000°C and 100 % humidity, resulting in a Si-SiO$_2$ substrate with a typical SiO$_2$-thickness of 600 nm. Afterwards Nb and Al are DC-magnetron sputtered *in-situ* from 3" targets of 99.999 % purity in argon (5.0) atmosphere at pressures of 0.96 and 0.72 Pa, respectively forming the Nb/Al-AlO$_x$/Nb multi-layer. Typical thicknesses for the ground electrode, Al-layer, and top electrode are 90 nm, 6 nm and 30-90 nm, respectively. Dry oxidation of the Al is performed in the load-lock of the system in pure oxygen (5.0).

Open squares in Fig. 1a show the dependence of the critical current density $j_c$(4.2 K) on oxidation atmosphere $p_{ox} \cdot t_{ox}$. The oxidation of the Al takes place in the load lock of the sputter

This work was supported by the Deutsche Forschungsgemeinschaft (via SFB/TRR-21 project A5) and the Center for Functional Nanostructures (CFN Project A 4.3).

J. M. Meckbach, M. Merker, S. J. Buehler K. Ilin and M. Siegel are with the Institute of Micro- and Nanoelectronic Systems – Karlsruhe Institute of Technology, Hertzstrasse 16, 76187 Karlsruhe, Germany (phone: +49-721-608-44995; fax: +49-721-757925; e-mail: m.meckbach@kit.edu)

B. Neumeier, U. Kienzle, E. Goldobin, R. Kleiner, D. Koelle are with the Physikalisches Institut–Center for Collective Quantum Phenomena and their Applications in LISA+, Universität Tübingen, Auf der Morgenstelle 14, 72076 Tübingen, Germany.

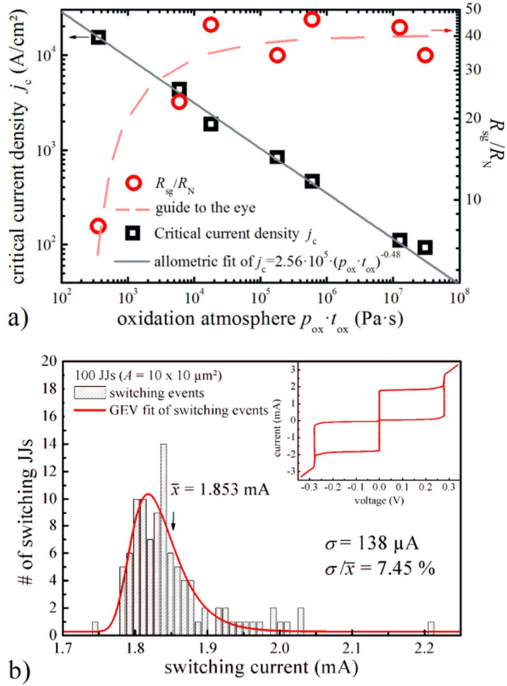

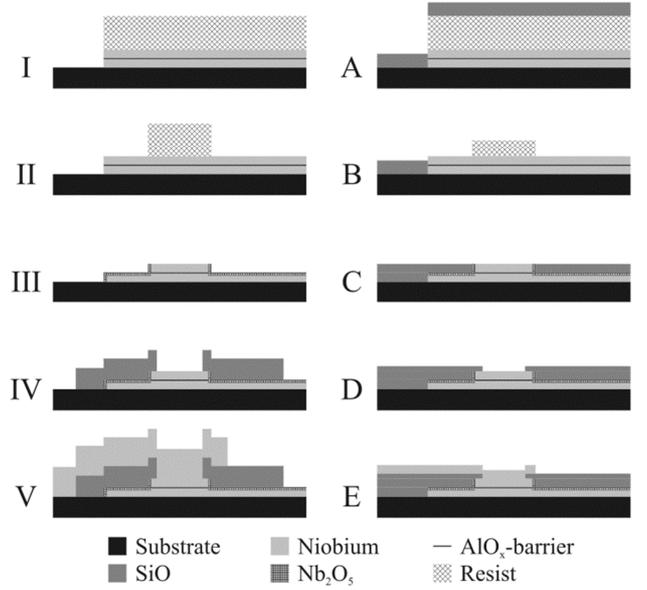

Fig. 2. Schematic representation of the conventional fabrication process (I-V) and the refined self-planarized process (A-E). The resulting step heights of the new process are approximately nine times less than of the conventional process.

Fig. 1. a) Open squares show the dependence of the critical current density $j_c$(4.2 K) of the fabricated tri-layers on the oxidation atmosphere. Open circles denote the $R_{sg}/R_N$-ratios of the selected trilayers. b) Histogram of 100 JJs of identical design connected in a serial array. The histogram is fitted with a generalized extreme value GEV distribution (solid red line). The inset shows the *IVC* of the measured array.

system. A reproducible $j_c(p_{ox} \cdot t_{ox})$-dependence is achieved using a controlled oxidation procedure resulting in critical current densities from $j_c = 50$ A/cm² - 15.3 kA/cm². For current densities up to $j_c = 2$ kA/cm² the oxidation time is kept at $t_{ox} = 30$ min and the oxidation pressure is varied between $p_{ox} = 10^6 - 2.5 \cdot 10^9$ Pa. For $j_c > 2$ kA/cm² the oxygen pressure $p_{ox}$ is kept at $10^6$ Pa and the oxidation time $t_{ox}$ is reduced. The critical-current densities achieved using this technique can be approximated by $j_c = 2.58 \cdot 10^5 \cdot (p_{ox} \cdot t_{ox})^{-0.48}$. This result is comparable to $j_c$-dependences found in [16]. Open circles in Fig. 1a depict the ratio of the sub-gap resistance $R_{sg}$ measured at 2 mV on the retrapping branch of the current-voltage-characteristic *IVC* and the normal-state resistance $R_N$ of the JJ. As can be seen the typical $R_{sg}/R_N$-ratio is above 30 and reaches up to 50 in some cases for $j_c < 2$ kA/cm². For higher $j_c$'s the $R_{sg}/R_N$-ratio decreases while still having reasonably good values of $R_{sg}/R_N = 8$ for current densities of 15.3 kA/cm².

The homogeneity of the fabricated trilayers has been checked by evaluation of *IVC*-measurements of serial arrays of JJs like shown in Fig. 1b (100-JJ-array). The measured $I_c$'s shown in the inset of Fig. 1b have not been corrected for variation of the effective JJ-size. Still, the coefficient of variation is merely 7.45 %. Such small deviations in an array over a length scale of ~2.5 mm reinforce the assumption of a homogeneous critical-current density over the entire chip. We should note that few JJs in the array deviate more strongly, which is comparable to results found in [17].

### B. Conventional Fabrication Process

Roman numerals in Fig. 2 label selected steps of the so-far used conventional fabrication process. The bottom electrode is patterned using positive photolithography and subsequent reactive-ion etching (RIE) of Nb using a $CF_4$-$O_2$-gas mixture, ion-beam etching (IBE) of the Al-$AlO_x$ barrier and again RIE (Fig. 2-I). Next, the resist stencil for the JJ definition is patterned (Fig. 2-II) and subsequently the top electrode is etched using RIE. The surface surrounding the protected JJ is anodized in an aqueous solution of $(NH_4)B_5O_8$ and $C_2H_6O_2$ (Fig. 2-III). Alternatively to the here shown procedure, a hard-mask process, as described in [14], can be employed. At this point the step height of the resulting chip topology is defined by the thickness of the bottom- / top-electrodes ($d_e \approx 90$ nm). Using a lift-off technique, SiO is thermally evaporated, insulating the JJ and the bottom electrode from the following wiring layer (Fig. 2-IV). To ensure a good edge-coverage the thickness is typically chosen to $d_{SiO} \approx 250 - 300$ nm. The final Nb wiring layer is again patterned using lift-off (Fig. 2-V). To overcome the SiO steps ($d_{SiO}$) typical thicknesses are in the range of $d_{wl} = 400 - 450$ nm.

Even though the existing technology was sufficient to fabricate sub-µm JJs, the large thicknesses of this conventional process strongly limit the minimal lateral sizes of all features. Especially for layers patterned in a lift-off technique the minimum width is typically in the range of thickness of the used resist, which in turn is usually twice as thick as the layer to be patterned. Consequently, the minimum width of the 450 nm thick wiring layer was limited to ~1 µm by default. Furthermore the resist thickness should be significantly larger than the step heights so that a homogeneous resist layer can be ensured. This made the use of thick resist even for electron-beam lithography (EBL) necessary again limiting the minimum feature size [14].

In order to overcome this limitation the layers need to be thinner, which presumes a topology with reduced step-heights. We refined the process to make it self-planarized, allowing sub-µm dimensions in all layers.

*C. Self-Planarized Fabrication Process*

Latin letters in Fig. 2 label selected steps of the newly refined self-planarized fabrication process. For sub-µm pattering EBL was employed with typical resist thicknesses of $d_{EBL} < 500$ nm. To minimize the large step heights of the above presented conventional process, we introduced self-aligned SiO layers after each etching process. In order to uphold an acceptable turn-around time of approximately 5 days for the entire fabrication process, a mix-&-match process based on negative resist (AR-N-7520.18 by AllResist©) was developed for the definition of large area bond pads and feed-lines using photolithography. This allows a fast standardized definition of the large area structures, while the small features may be defined subsequently using EBL. For a sufficient stability of the resist with respect to the various etching steps and the following lift-off procedure a thorough optimization of the lithography and baking parameters was necessary. We found that the resist becomes highly stable with respect to plasma etching (IBE and RIE) with ~10 times increased pre-baking times (10 min) and an EBL exposure dose reduced by a factor of ~8 (24 µC/cm²) as compared to the recommendations of the manufacturer.

The planarization of the trilayer is performed in two steps where the SiO layer planarizing the ground electrode is deposited after the initial etching of the trilayer with a thickness $d_A = d_{ge} + \Delta$ (see Fig. 2-A and 2-B). The additional $\Delta$ is determined by the selectivity of the etching rates between Nb and SiO in the following etching process, for the definition of top electrode and junction. After the JJ-definition the first planarization SiO is equally thick as the bottom electrode, so that a second SiO layer deposited after the anodic oxidation fully levels the surface of the structure as can be seen in Fig. 2-C. The characteristic step height after JJ definition is in the range $d_{step} < 10$ nm, which is an improvement of approximately a factor 9 as compared to the before described conventional process.

Due to the isotropic nature of RIE an overhang of the resist stencil over the etched Nb is created. Later thermally evaporated SiO will result in a poor edge coverage around the JJ-definition due to the anisotropic nature of the deposition (see schematic in Fig. 3). Such trenches can be minimized to some degree using isotropically sputtered isolation materials. Alternatively reactive-ion etching may be replaced with ion-beam etching, exhibiting a much more anisotropic etching profile. Using this process we achieved trenches as narrow as 10 nm. In order to uphold a high yield, an additional isolation layer covering the mentioned trenches is advisable (see Fig. 2-D). The thickness of this layer is typically in the range of $d_D = 30$ nm and defines the maximum step height of the topology.

Finally the top wiring layer is deposited, having a typical thickness of 150-200 nm (see Fig. 2-E). If additional wiring layers are needed the layer is again pattered in a standard etching process using a preliminary deposited AlN layer as an etch stop and subsequently planarized using AlN or SiO. When comparing the final cross section of the conventional process (Fig. 2-V) with that of the self-planarized process (Fig. 2-E) one notices that the maximum step height even with non-planarized wiring layer is approximately 50 % of the conventional process and that moreover much less steps are created.

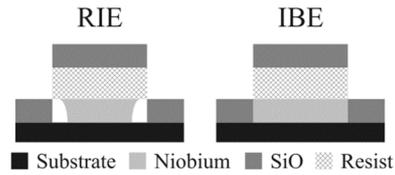

Fig. 3. Schematic representation of a self-aligned deposition of SiO after RIE (left) and IBE (right). Due to the isotropic nature of RIE, trenches around the JJ-definition will form which may lead to shortcuts at a later stage of the fabrication process.

## III. MEASUREMENTS

All measurements were performed at $T = 4.2$ K in a liquid-helium transport dewar. Currents were supplied by battery-powered current sources. The voltage amplifiers were also battery powered in order to minimize fluctuation during the four-point measurements.

Typical values of the characteristic voltage $V_m = I_c R_{sg}$ for a single unshunted junction are comparable to values from literature [15] and range from 20 to 80 mV at $T = 4.2$ K. Employing the newly refined process we were able to fabricate junctions with significantly reduced lateral dimensions. Fig. 4a shows the *IVC* of a JJ with an area of 0.25 µm². Despite the very small feature size the quality parameters are extraordinarily good with a gap-voltage $V_{gap} = 2.81$ mV and $R_{sg}/R_N = 37$. Fig. 4b shows an SEM image of a comparable junction with an approximate area of 0.4 x 0.4 µm². Such junctions are ideal candidates for the use as SIS mixer elements [11].

Besides the investigation of small JJs, we are currently investigating long Josephson junctions (LJJs) for the use as Flux-Flow Oscillators (FFOs) [18] or investigations of artificial vortices [4,5,14]. The latter requires LJJs which are significantly longer than the effective Josephson penetration depth $L_{JJ} >> \lambda_{J,eff}$ while the width is ideally in the sub-µm range $W_{JJ} << \lambda_{J,eff}$ [19]. As proposed in [4,5] creation of fractional vortices or so-called $\kappa$-vortices [2-5,20] can be achieved by injection and extraction of a dc-injector current $I_{inj}$ through tiny feed lines ($w_{inj} << \lambda_{J,eff}$) separated by $dX \approx W_{JJ}$. Various devices based on such injector designs are conceivable including qubits [12] or meta-materials [15]. Both of these highly sophisticated devices require sub-µm injector widths $w_{inj}$ due to the restrictions given by the small Josephson penetration depth resulting from the necessity of high critical-current densities.

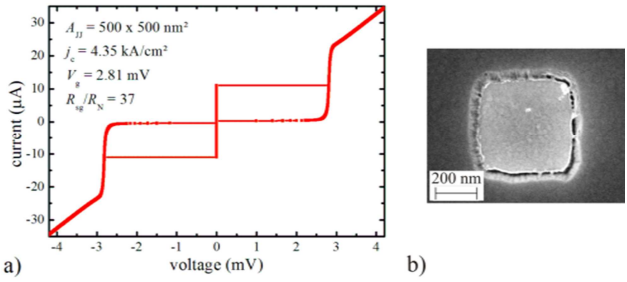

Fig. 4. a) *IVC* of a single unshunted sub-µm JJ with a designed area of 0.25 µm² measured at $T$ = 4.2 K. b) SEM image of a junction definition of the area 0.16 µm².

So far, sub-µm patterning in the top wiring layer was only possible when employing CMP techniques. Using our newly developed process we were able to fabricate first prototype devices of long JJs with sub-µm junction width $W_{JJ}$ and injector width $w_{inj}$. Fig. 5a shows the dependence of the critical current $I_c$ on the applied dc-injector current of the LJJ shown in the corresponding SEM image in Fig. 5b. With $L_{JJ} = 40$ µm, $W_{JJ} = 0.83$ µm and an effective Josephson penetration depth $\lambda_{J,eff} \approx 8.6$ µm ($j_c \approx 3$ kA/cm²) the LJJ is sufficiently long to study Josephson vortex behavior. Also the requirements $w_{inj} = 0.6$ µm $<< \lambda_{J,eff}$ and $dX = 0.3$ µm $<< \lambda_{J,eff}$ are met. $I_c(I_{inj})$-dependencies are numerically analyzed in [20], and the amplitude of the side maxima could be roughly estimated to 1-$I_c(\kappa=\pm 2\pi)/I_c(\kappa=0) \approx (w_{inj} + dX)/L_{JJ}$. For a design as shown in Fig. 5b this decrease is expected to be less than 2 %. In the $I_c(I_{inj})$-dependence no effective decrease in the amplitude could be measured proving the high quality of the fabricated devices. Further investigations of $I_c(I_{inj})$-dependence and of the macroscopic quantum-tunneling (MQT) behavior on the injector current is ongoing and will be presented elsewhere.

## IV. CONCLUSION

We have refined a fabrication process for Josephson junctions in overlap geometry with regards to the surface self-planarization. The final step height may be reduced to a few 10 nm even without CMP procedures and thus allowing the use of electron-beam lithography (EBL) even in the top wiring layer. Well-adjusted parameters for the different lithography steps make so-called mix-&-match lithography possible combining photolithography and EBL using the same resist. The resulting resist stencils are strong enough to withstand reactive-ion etching as well as ion-beam etching and a subsequent self-aligned deposition of SiO while still being well dissolved in acetone for lift-off afterwards. Elaborate series of transport measurements show that the process yields very high quality parameters and a good reproducibility for junctions down to a few 100 nm for $j_c$ ranging from 50 A/cm² - 15 kA/cm². The smallest junctions measured had a designed area of 0.023 µm² with an $I_c \approx 2.45$ µA at $T = 4.2$ K. LJJs with sub-µm current injectors have also been fabricated and first characterizations of the dependence of the critical current $I_c$ on the dc-injector current $I_{inj}$ shows reasonable agreement with theory.

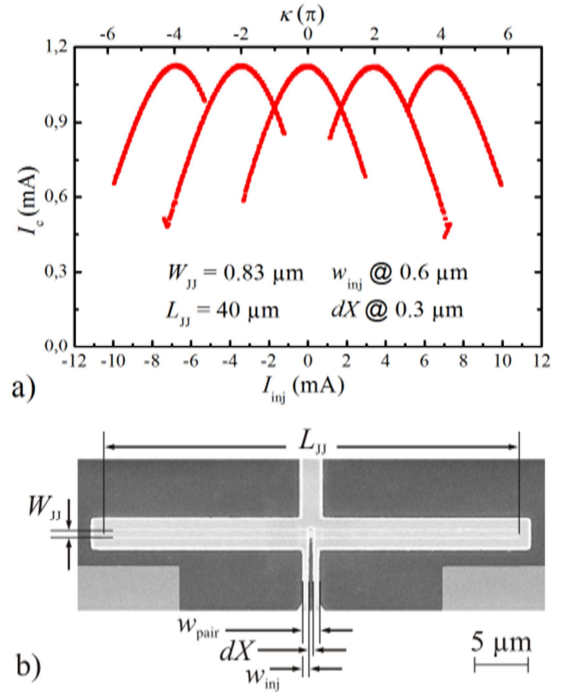

Fig. 5. a) $I_c(I_{inj})$-dependence of the LJJ shown in b) measured at $T$ = 4.2K. b) SEM image of a LJJ of sub-µm width and nano injectors.

## V. ACKNOWLEDGEMENT

This work was supported by the Deutsche Forschungsgemeinschaft (via SFB/TRR-21 project A5) and the Center for Functional Nanostructures (CFN Project A 4.3). J. M. Meckbach would like to thank the IEEE CSC for granting the *2011 IEEE CSC graduate student fellowship award*.